\documentclass[aps,twocolumn,showpacs,preprintnumbers,amsmath,amssymb,linenumbers,prb]{revtex4}
\usepackage{graphicx}
\usepackage{dcolumn}
\usepackage{bm}
\usepackage{braket}
\usepackage{color}

\begin{document}

\title{Exact-diagonalization study of exciton condensation in electron bilayers}

\author{T. Kaneko$^1$, S. Ejima$^2$, H. Fehske$^2$, and Y. Ohta$^1$}
\affiliation{$^1$Department of Physics, Chiba University, Chiba 263-8522, Japan\\
$^2$Institut f\"ur Physik, Ernst-Moritz-Arndt-Universit\"at Greifswald, 17489 Greifswald, Germany}

\date{\today}

\begin{abstract}
We report on small-cluster exact-diagonalization calculations which prove the formation of electron-hole pairs (excitons) 
as prerequisite for spontaneous interlayer phase coherence in bilayer systems described by the extended Falicov-Kimball model.  
Evaluating the anomalous Green's function and momentum distribution function of the pairs, and thereby analyzing the 
dependence of the exciton binding energy, condensation amplitude, and coherence length on the Coulomb interaction strength, 
we demonstrate a crossover between a BCS-like electron-hole pairing transition and a Bose-Einstein condensation of tightly 
bound preformed excitons.  We furthermore show that a mass imbalance between electrons and holes tends to suppress the 
condensation of excitons.  
\end{abstract}


\maketitle

\section{Introduction}

The formation of excitonic quantum condensates is an intensively studied 
continuous problem in condensed matter physics.\cite{BBB62,KK65,MS00,LEKMSS04}
In a two-component (electron-hole) many-particle system the attractive Coulomb interaction 
between oppositely charged electrons and holes can trigger their pairing and -- under 
certain conditions -- build up a macroscopic phase-coherent quantum state.  

A variety of experimental attempts have been made to observe the condensed state of excitons in 
quasi-thermal equilibrium, e.g., in photoexcited semiconductors such as 
Cu$_2$O,\cite{SWM90,LW93,NN05,YCK11,SSKSKSSNKF12} or in unconventional semiconductor 
and bilayer graphene systems subject to electric and/or magnetic fields.\cite{BGC02,SDLPW02,EM04,CGPNG09,NL10}  
Quite recently, the emergence of spontaneous coherence in a gas of indirect excitons 
in an electrostatic trap has been reported.\cite{HLRBHC12}  
Neutral electron-ion quantum plasmas are other promising candidates for exciton 
condensates.\cite{FFBFL07,SFBF12} 

From a theoretical point of view, a possible continuous transition between a Bardeen-Cooper-Schrieffer 
(BCS) electron-hole pair condensate and a Bose-Einstein condensate (BEC) of preformed excitons 
has been of topical interest.\cite{LEKMSS04,KK68,CN82,NC82,NS85,MRR90,PNS07}  However, exact results for 
the ground-state properties of strongly correlated electron-hole (excitonic) systems are rare.  
Gas (or fluid) models have recently been studied, e.g., by the diffusion quantum Monte Carlo 
method.\cite{DRS02,MRNO13}  Lattice fermion models with short-range Coulomb interaction, such as 
multi-band Hubbard-like models,\cite{THO06,OT09,TO10} should be capable of describing the physics of 
exciton condensation as well, but have not yet been thoroughly explored by unbiased numerical techniques.  

Motivated by this situation, in this paper, we made an attempt to address the problem of 
exciton condensation in electron-hole bilayers in terms of a minimal lattice fermion model,
the so-called extended Falicov-Kimball model (EFKM).\cite{FK69,Ba02b,Fa08,SC08,IPBBF08} 
Originally the EFKM describes  a two-band electron system with local Coulomb interaction 
between $f$- and $c$-band electrons and has been used  to study electronic 
ferroelectricity,\cite{Ba02b,BGBL04,Fa08} excitonic resonances,\cite{PFB11} 
or the excitonic insulator state.\cite{IPBBF08,ZIBF10,ZIBF11,ZIBF12,SEO11,AK12}  
Different from  these problems, in our double-layer (DL) system, the numbers of $f$- and $c$-particles 
are separately conserved, however, because charge transfer between the two layers is assumed to 
be impossible.  This rather mimics the generic situation in semiconductor electron-hole double 
quantum wells,\cite{EM04,ZLHR95,SMRL09} bilayer quantum antiferromagnets,\cite{RWHZ12} and 
double-monolayer\cite{MBSM08,PF12} or double-bilayer graphene.\cite{PNH13}  

\section{Model}

\begin{figure}[t]
\begin{center}
\includegraphics[width=\linewidth]{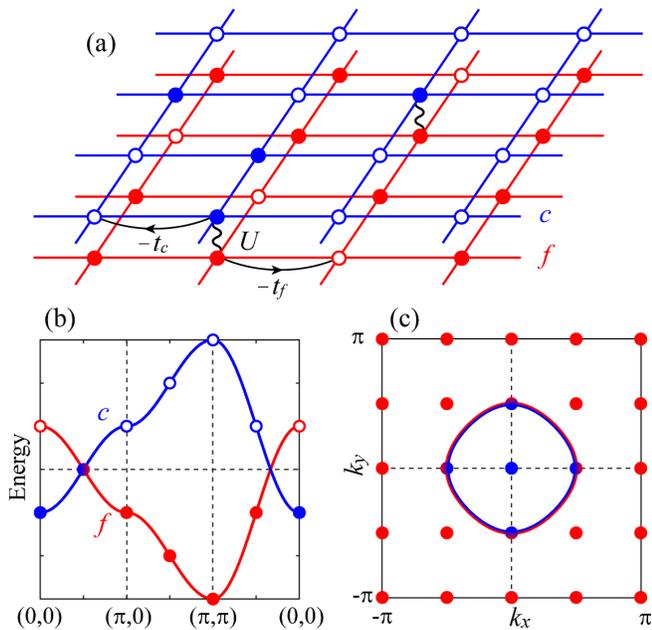}
\caption{(Color online) (a) Schematic representation of the DL EFKM cluster model with 
$N_s=16$ sites (32 orbitals).  (b) Non-interacting tight-binding band structure and (c) square lattice Brillouin zone. 
Dots indicate the allowed momenta of the $4\times4$ lattice with periodic 
boundary conditions.  Throughout this work, 
we assume filling factors $n^f=0.75$ and $n^c=0.25$, i.e., $(N_f,N_c)=(12,4)$ which means $n^h=n^e=0.25$,  
irrespective of $U$.  The red and blue lines in (c) show the perfectly  matching hole and electron Fermi surfaces, respectively,
with  finite-lattice Fermi momenta $\bm{k}_{\rm F}$  located at $\bm{k}=(\pm \pi/2,0)$ and $(0,\pm \pi/2)$. 
}\label{fig1}
\end{center}
\end{figure}

The EFKM for an electron-hole  DL takes the form 
\begin{align}
\mathcal{H}=-&t_{f}\sum_{\langle i,j\rangle}(f^{\dag}_{i}f_{j}^{}+{\rm H.c.})
-t_{c} \sum_{\langle i,j\rangle}(c^{\dag}_{i} c_{j}^{}+{\rm H.c.}) \notag \\
-&\mu_f \sum_{i}n_i^f-\mu_c\sum_{i}n_i^c+U\sum_{i}n_i^fn_i^c, 
\label{ham}
\end{align} 
where $f^{\dag}_{i}$ ($f_{i}^{}$) creates (annihilates)  an electron in the $f$-orbital at site $i$ of the hole 
(or valence-band) layer, and $n_i^f=f^{\dag}_{i}f_{i}^{}$ is the $f$-particle number operator.  
The transfer amplitude between $f$-orbitals on nearest-neighbor sites is denoted by $t_f$.  
Corresponding definitions apply for the $c$-orbital of the electron (or conduction-band) layer.  
$U$ $(>0)$ parametrizes the on-site interlayer (on-site) Coulomb attraction between $f$-holes and $c$-electrons.  
The spin degrees of freedom have been ignored for simplicity.  
Furthermore we assume a band structure with a direct band gap ($t_c\cdot t_{f}< 0$) as shown in Fig.~\ref{fig1}.  

Taking into account the experimental situation,\cite{SWM90,LW93,NN05,YCK11,SSKSKSSNKF12,BGC02,SDLPW02,EM04,CGPNG09,PF12,SFBF12} 
we assume that the excited electrons and holes have infinite lifetime, that the number of excited electrons 
is equal to the number of excited holes, and that the number of bound pairs (excitons) can be viewed as an input parameter, 
independent of the interaction strength.  
In practice, we adjust the chemical potentials $\mu_f$ and $\mu_c$ to maintain the number of electrons 
in the $f$- and $c$-layer separately, thereby fixing the average $f$- and $c$-particle density per site 
as $n^f$ and $n^c$, respectively.  Due to this simplified description, issues such as exciton Mott transition 
and biexciton formation\cite{FFBFL07} are beyond the scope of this work.  

\section{Theoretical Approach}

We employ a Lanczos exact-diagonalization technique for a finite square lattice with 
periodic boundary conditions (see Fig.~\ref{fig1}) and calculate the anomalous Green's function 
for exciton condensation 
\begin{align}
G^{cf}(\bm{k},\omega)=\Braket{\psi_0^N|c^{\dag}_{\bm{k}} 
\frac{1}{\omega+i0^+-\mathcal{H}+E_0}f_{\bm{k}}|\psi_0^N} 
\end{align}
in the momentum ($\bf k$) and frequency ($\omega$) space, 
where $|\psi_0^N\rangle$ is the ground-state wave function and $E_0$ is the ground-state 
energy of a system with $N$ electrons.  We define the anomalous spectral function 
\begin{align}
F({\bm k},\omega)=-\frac{1}{\pi}\Im G^{cf}(\bm{k},\omega)
\end{align} 
and denote its frequency integral 
by $F_{\bm k}$.  Clearly, the anomalous Green's function vanishes in finite 
systems without long-range phase coherence.  We therefore have to assume the presence 
of the state $|\psi_0^N\rangle$, which is a coherent superposition of states with different 
numbers of excited electrons and holes (or excitons) at given number $N$, just as for 
the BCS wave function of superconductors where the number of electrons is also not conserved.  
In order to detect particle fluctuations of the exciton condensate, we adopt a technique 
introduced for the  evaluation of the superconducting anomalous Green's function on small 
clusters,\cite{OSEM94,ONETM95} which allows for the calculation of the off-diagonal Green's 
functions with respect to varying particle numbers [see Eq.~(\ref{Fk})].  
We thus monitor the excitonic pairing instability via the anomalous excitation spectrum 
(corresponding to the Bogoliubov quasiparticle spectrum in superconductors).  
Note that the term `anomalous' is used to indicate that the number of electrons on each 
of the $f$- and $c$-bands is not conserved in the course of exciton condensation (or 
spontaneous $c$-$f$ hybridization) although the total number of electrons $N$ is conserved.  

Having $G^{cf}(\bm{k},\omega)$ determined, we can calculate the condensation amplitude 
$F_{\bm{k}}$ (following Refs.~\onlinecite{OSEM94} and \onlinecite{ONETM95}) from  
\begin{align}
F_{\bm{k}}=\Braket{N_f-1,N_c+1|c^{\dag}_{\bm{k}}f_{\bm{k}}^{}|N_f,N_c}, 
\label{Fk}
\end{align}
where $\ket{N_f,N_c}$ is the ground state with the fixed numbers of $f$- and $c$-electrons,  
and subsequently will be able to determine the order parameter  
\begin{align}
\Delta = \frac{U}{N_s}\sum_{\bm{k}} F_{\bm{k}}
\end{align}
and the coherence length 
\begin{align}
\xi = \sqrt{\frac{\sum_{\bm{k}}|\bm{\nabla}_{\bm{k}} F_{\bm{k}}|^{2}}{\sum_{\bm{k}}|F_{\bm{k}}|^{2}}} 
\end{align}
for the excitonic condensate ($N_s$ counts the number of lattice sites).  

\begin{figure*}[t]
\begin{center}
\includegraphics[width=\linewidth]{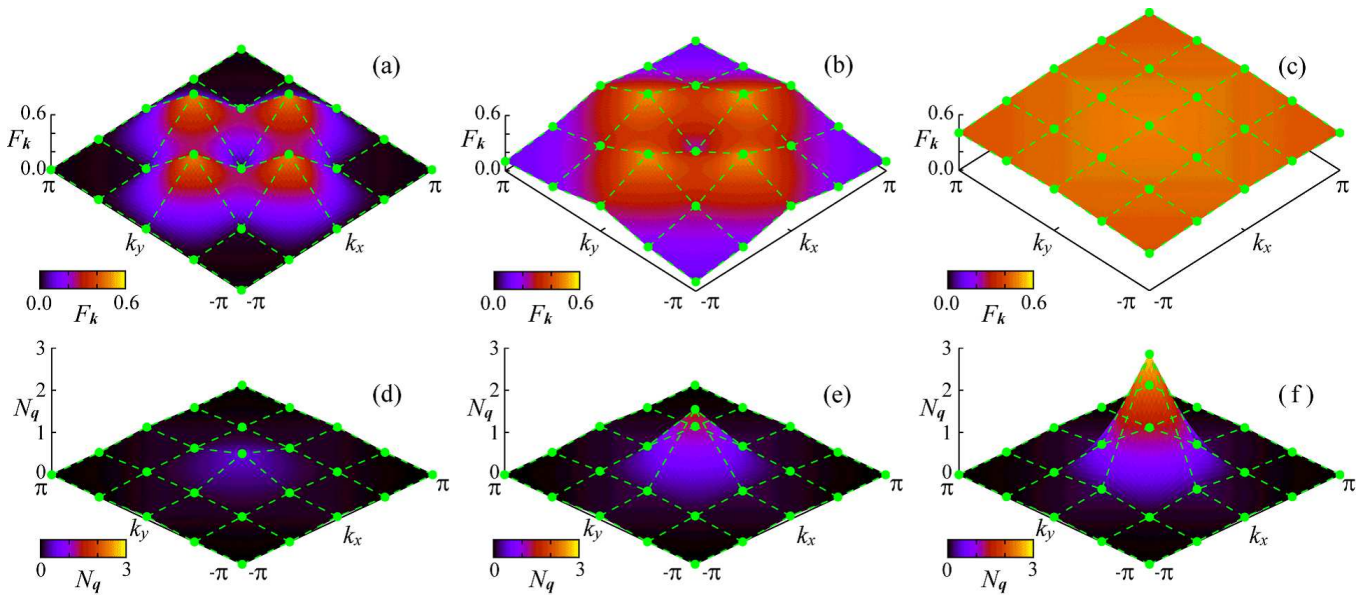}
\caption{(Color online) Condensation amplitude 
$F_{\bm{k}}$ (upper panels) and momentum distribution 
function  $N_{\bm{q}}$ of excitons (lower panels) in the mass-symmetric DL EFKM with $U/t=0.5$ (left), 5 (middle), and 50 (right).  
}\label{fig2}
\end{center}
\end{figure*}

The binding energy of an exciton $E_{B}$ should be equal to twice of the order parameter $\Delta$ 
in the weak-coupling limit and deviate largely from this value in the strong-coupling regime.  
Within our finite-cluster approach, $E_B$ may be obtained representing the orbital flavor by 
electron-hole variables, i.e., $f^{\dag}_i\rightarrow h_{i}$ and $c^{\dag}_{i}\rightarrow e^{\dag}_i$.  
As a result, the interaction term of the DL EFKM  takes the form 
$U\sum_{i}n_i^fn_i^c\rightarrow -U\sum_{i} n_i^en_i^h+U\sum_{i}n_i^e$, 
where, in addition to the attractive electron-hole interaction, an extra on-site energy term 
appears.  Due to this term, we should first determine the energy for the addition and 
removal of an electron: 
\begin{align}
E^{+}_{B}&=E_0(N_f-1,N_c+1)+E_0(N_f,N_c)\nonumber\\&\quad-2E_0(N_f,N_c+1) + U\,, \\
E^{-}_{B}&=E_0(N_f-1,N_c+1)+E_0(N_f,N_c)\nonumber\\&\quad-2E_0(N_f-1,N_c) - U\,, 
\end{align}
where $E_0(N_f,N_c)$ is the ground-state energy of the system with $(N_f,N_c)$ electrons.  
Then, if $|t_f|=t_c$, the exciton binding energy $E_B$ equals $E^{+}_{B}=E^{-}_{B}$.  
For the mass-asymmetric case $|t_f|\ne t_c$, however, $E^{+}_{B}\ne E^{-}_{B}$ 
because $E_0(N_f,N_c+1) -U \ne E_0(N_f-1,N_c)$.  
Hence, $E_B$ should be defined as the average of $E^+_B$ and $E^-_B$, i.e., in general 
the exciton binding energy is given by
\begin{align}
E_{B}=&E_0(N_f-1,N_c+1)+E_0(N_f,N_c) \notag \\
      &-E_0(N_f-1,N_c)-E_0(N_f,N_c+1)\,.
\end{align}

Finally, introducing a creation operator $b^{\dag}_{\bm{q}}=(1/\sqrt{N_s})\sum_{\bm{k}}
c^{\dag}_{\bm{k}+\bm{q}}f_{\bm{k}}^{\phantom{\dagger}}$ of an excitonic quasiparticle 
with momentum $\bm{q}$, the momentum distribution function of excitons 
can be obtained from  
\begin{align}
N_{\bm{q}}= \Braket{N_f,N_c|b^{\dag}_{\bm{q}}b_{\bm{q}}^{} |N_f,N_c}\,. 
\end{align}

\section{Numerical Results}

\subsection{Mass-symmetric case}

We now present the results of our exact-diagonalization study.  
Let us first examine the DL EFKM without mass imbalance, i.e., $|t_f|=t_c\equiv t$.  
Figure~\ref{fig2} shows the corresponding data for the condensation amplitude $F_{\bm{k}}$ 
and the exciton momentum distribution $N_{\bm{q}}$, in a wide parameter range of $U/t$.  
In the weak-coupling regime [panels (a) and (d)], $F_{\bm{k}}$ exhibits pronounced 
maxima at the Fermi momenta $\bm{k}_{\rm F}=(\pm\pi/2,0)$, $(0,\pm\pi/2)$ and decreases 
rapidly away from the `Fermi surface', pointing towards a BCS-type instability of weakly 
bound electron-hole pairs with $s$-wave symmetry.  
As $U/t$ increases, $F_{\bm{k}}$ broadens in momentum space [panel (b)], indicating that 
the radius of the bound electron-hole objects becomes smaller in real space.  
Accordingly, $N_{\bm{q}}$ is enhanced at momentum $\bm{q}=(0,0)$; see Fig.~\ref{fig2} (e).  
In the strong-coupling regime [panels (c) and (f)], $F_{\bm{k}}$ is homogeneously spread 
over the entire Brillouin zone, whereas $N_{\bm{q}}$ is sharply peaked at $\bm{q}=(0,0)$, 
which is a sign of a BEC of tightly bound excitons.  That is to say, as the attraction 
between electrons and holes increases in the DL EFKM, we get evidence for a BCS-BEC 
crossover scenario.

\begin{figure}[t]
\begin{center}
\includegraphics[width=\linewidth]{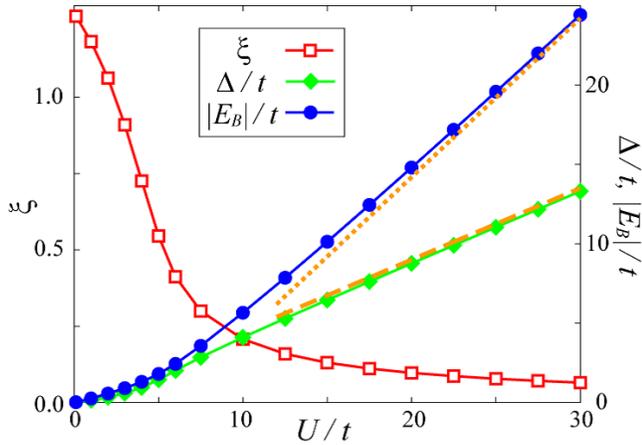}\\
\caption{(Color online) Coherence length $\xi$ (squares), order parameter $\Delta$ (diamonds), 
and exciton binding energy $E_B$ (circles) for the mass-symmetric DL EFKM as functions of $U/t$.  
For comparison, the asymptotics in the  strong-coupling limit $\Delta \propto 0.45 U$ (dashed line) 
and $|E_B| \propto U$ (dotted line) have been inserted.
}\label{fig3}
\end{center}
\end{figure}

The behavior of the coherence length depicted in Fig.~\ref{fig3} as a function of the Coulomb 
attraction corroborates this finding.  The spatial coherence of the excitonic state decreases 
with increasing $U/t$, indicating that the character of the condensate changes from BCS-like 
to BEC-like.  That $\xi$ stays finite as $U/t\to 0$ is an obvious artifact  of our small cluster calculation.  
Figure~\ref{fig3} also displays the functional dependence of both the exciton order parameter 
and the exciton binding energy on $U/t$.  The results may be compared with those of the BCS 
mean-field theory,\cite{NS85,MRR90} which gives $\Delta$ and $E_B$ as solution of the 
self-consistent equations 
\begin{align}
1&=\frac{U}{2N_s}\sum_{\bm{k}}\frac{1}{{\sqrt{(\varepsilon_{\bm{k}}-\bar{\mu})^{2}+\Delta^2}}}\,,\\ 
2n&=1-\frac{1}{N_s}\sum_{\bm{k}}\frac{\varepsilon_{\bm{k}}-\bar{\mu}}{{\sqrt{(\varepsilon_{\bm{k}}-\bar{\mu})^{2}+\Delta^2}}}\,, 
\end{align}
where $\varepsilon_{\bm{k}}=2t(\cos k_x+\cos k_y)$, $n=n^e=n^h$, 
$\bar{\mu}=\mu-U(n-1/2)$, and $\mu_f=-\mu_c=\mu$.  

In the weak-coupling limit, we should recover the usual BCS picture.  
$\Delta$ should therefore increase exponentially with 
$U$: $\Delta\propto\exp(-1/\rho(\varepsilon_{\rm F})U)$, thereby 
satisfying the relation $|E_B|=2\Delta$ with $\rho(\varepsilon_{\rm F})$ 
being the density of states at the Fermi level.  
In the strong-coupling limit, on the other hand, the BCS equations 
yield the asymptotic behavior: 
$\Delta=U\sqrt{n(1-n)}=\sqrt{3}U/4\simeq 0.433U$ and 
$|E_{B}| = 2\sqrt{\bar{\mu}^{2}+\Delta^{2}} = U $.  
The numerical results obtained for $\Delta$ and $|E_B|$ show that 
we find the BCS relation $|E_B|=2\Delta$ at weak couplings.  
In the strong-coupling limit, $\Delta$ and $|E_B|$ are found to be 
$\propto 0.45U$ and $\propto U$, respectively, which matches the 
BEC of composite bosons, where $\Delta=0.433U$ and $|E_B|=U$ 
for $U/t\rightarrow\infty$.  

\subsection{Mass-asymmetric case}
\begin{figure}[t]
\begin{center}
\includegraphics[width=\linewidth]{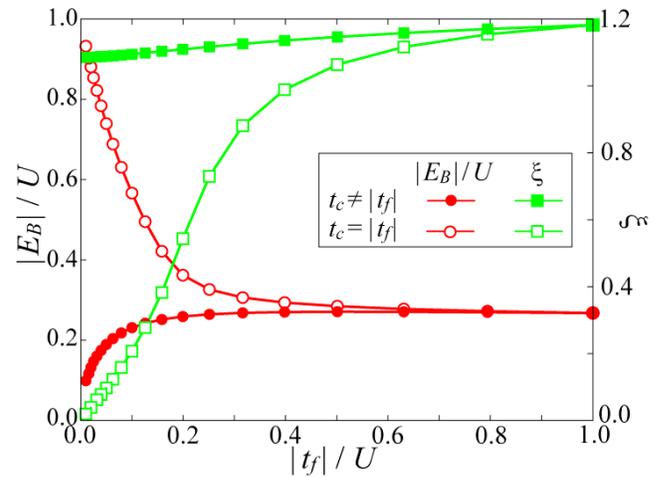}
\caption{(Color online) Binding energy $E_B/U$ (left ordinate) 
and coherence length $\xi$ (right ordinate) for the mass-asymmetric 
(filled symbols) and mass-symmetric (open symbols) DL EFKM as functions 
of $|t_f|/U$ at $t_c/U=1$.   
}\label{fig4}
\end{center}
\end{figure}

We finally address the effects of a mass imbalance between $f$ holes and $c$ electrons.  
Since $|t_f|\ne t_c$ it makes sense to use $U$ as the unit of energy and determine 
the exciton binding energy $E_B$ and coherence length $\xi$ in dependence on $|t_f|/U$.  
Figure~\ref{fig4} shows the results for $t_c/U=1$ in comparison to the mass-symmetric 
case where a BCS-to-BEC transition occurs with decreasing $|t_f|/U$.  By contrast, 
$\xi$ is not reflective of such a crossover for $t_c\ne|t_f|$, and the exciton 
binding energy even weakens at strong couplings $|t_f|/U\ll 1$.  

In the strong-coupling region, where both $|t_f|/U$ and $t_c/U$ are small, 
the EFKM can be mapped onto the XXZ quantum spin-1/2 model in a magnetic 
field,\cite{Ba02b}   
\begin{align}
\mathcal{H}_{\rm eff}=J\sum_{\langle i,j\rangle}[\bm{\tau}_i\cdot\bm{\tau}_j+\delta\tau^z_i\tau^z_j]-B_z\sum_{i}\tau^z_i
\end{align}
with $\bm{\tau}_i=(1/2)\sum_{\alpha,\beta}\alpha^{\dag}_i{\bm \sigma}_{\alpha\beta}\beta_i$ ($\alpha,\beta=f,c$;
${\bm \sigma}$ is the vector of Pauli matrices), 
$J = 4|t_f|t_c/U$, and $\delta=(|t_f|-t_c)^{2}/(2|t_f|t_c)$.  
$B_z=2\mu$ is determined in order to maintain $\sum_i\tau^z_i=1/4$.  
The effective model is isotropic in spin space for the case of $|t_f|=t_c$,  
and exhibits antiferromagnetic order in the $x$-$y$ plane at zero temperature.  
This long-range ordered state corresponds to an exciton condensate in the 
original EFKM.  Different  hopping parameters $t_c\ne|t_f|$ give rise to an 
Ising anisotropy $\delta$, which tends to suppress the $x$-$y$ antiferromagnetic 
order.  Accordingly, the exciton binding energy $|E_B|$ (excitonic condensate) 
is suppressed as $|t_f|/U\to 0$.  

\begin{figure}[t!]
\begin{center}
\includegraphics[width=\linewidth]{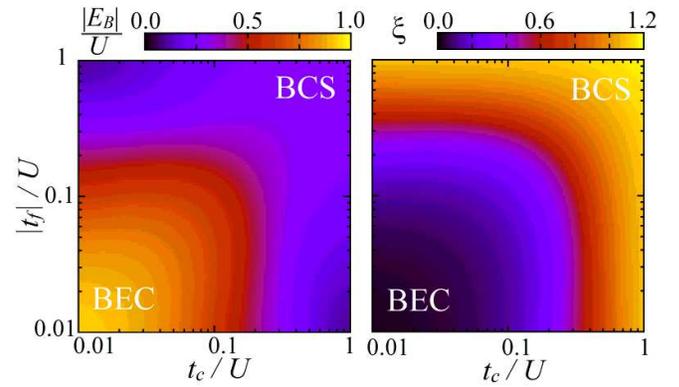}
\caption{(Color online) Exciton binding energy $E_B/U$ (left panel) and 
coherence length $\xi$ (right panel) of the DL EFKM in the $t_c/U$--$|t_f|/U$ plane.  
}\label{fig5}
\end{center}
\end{figure}

Figure~\ref{fig5} compiles our $E_B$ (left panel) and $\xi$ (right panel) data 
by two contour plots in the $t_c/U$-$|t_f|/U$ plane.  For the mass-symmetric case 
$t_c=|t_f|$, i.e., on the diagonals of Fig.~\ref{fig5}, both $|E_B|/U$ and $\xi$ 
indicate a smooth crossover from BCS to BEC as $U$ increases.  On the other hand, 
at sufficiently weak Coulomb interactions, $t_c/U\agt 0.3$, we stay in the 
BCS-like state as $|t_f|/U$ is varied by changing the absolute value of $t_f/t_c$.  
Note that a strong mass imbalance between electrons and holes acts in a 
`pair-breaking' way in both the BCS\cite{BF06} and BEC\cite{MIUA05,THO06} limits.

\section{Summary}

To give a r\'{e}sum\'{e}, based on unbiased exact-diagonalization calculations
for the two-dimensional extended Falicov-Kimball model, we have studied the 
formation of excitons in both mass-symmetric and mass-asymmetric electron-hole 
double-layer systems (bilayers) and provided, most notably, strong evidence for exciton 
condensation and a BCS-BEC crossover scenario at zero temperature.  Thereby, 
the properties of the excitonic quasiparticles and the nature of the condensation 
process were analyzed, exploiting the anomalous Green's function in order to 
determine the order parameter of the condensate and coherence length, as well as 
the binding energy and momentum distribution function of excitons.  
The weak and strong correlation limits are discussed and put into perspective to 
approximative analytical approaches.  
We corroborated previous analytical\cite{MIUA05,THO06,BF06} and numerical\cite{SFBF12} 
findings to that effect that a mass imbalance between electrons and holes might 
suppress the condensation of excitons.  This holds even in the strong coupling regime.  
We hope that the presented results will stimulate further experimental studies of 
exciton condensation in bilayer systems with strong electronic correlations.  

\begin{acknowledgments}
The authors would like to thank K. Seki, Y. Tomio, and B. Zenker 
for enlightening discussions.  This work was supported in part by the 
Kakenhi Grant No.~22540363 of Japan and by the Deutsche Forschungsgemeinschaft through SFB 
652 B5.  Part  of the computations have been  carried 
out at the Research Center for Computational Science, 
Okazaki Research Facilities, Japan.  
\end{acknowledgments}

\bibliography{ref}
\bibliographystyle{apsrev}

\end{document}